# Stochastic modeling of a serial killer

M.V. Simkin and V.P. Roychowdhury
Department of Electrical Engineering, University of California, Los Angeles, CA 90095-1594

We analyze the time pattern of the activity of a serial killer, who during twelve years had murdered 53 people. The plot of the cumulative number of murders as a function of time is of “Devil’s staircase” type. The distribution of the intervals between murders (step length) follows a power law with the exponent of 1.4. We propose a model according to which the serial killer commits murders when neuronal excitation in his brain exceeds certain threshold. We model this neural activity as a branching process, which in turn is approximated by a random walk. As the distribution of the random walk return times is a power law with the exponent 1.5, the distribution of the inter-murder intervals is thus explained. We illustrate analytical results by numerical simulation. Time pattern activity data from two other serial killers further substantiate our analysis.

Figure 1 shows a time-plot of the cumulative number of murders committed by Andrei Chikatilo [1] during his twelve-year activity. It is highly irregular with long time intervals without murder interrupted by jumps, when he murdered many people during a short period. Such a curve is known in mathematics as a “Devil’s staircase” [2]. We can characterize the staircase by the distributions of step lengths. Figure 2 shows such distributions for the staircase of Figure 1 in log-log coordinates. A linear fit shows that the exponent of the power law of the probability density distribution (in the region of more than 16 days) is 1.4.

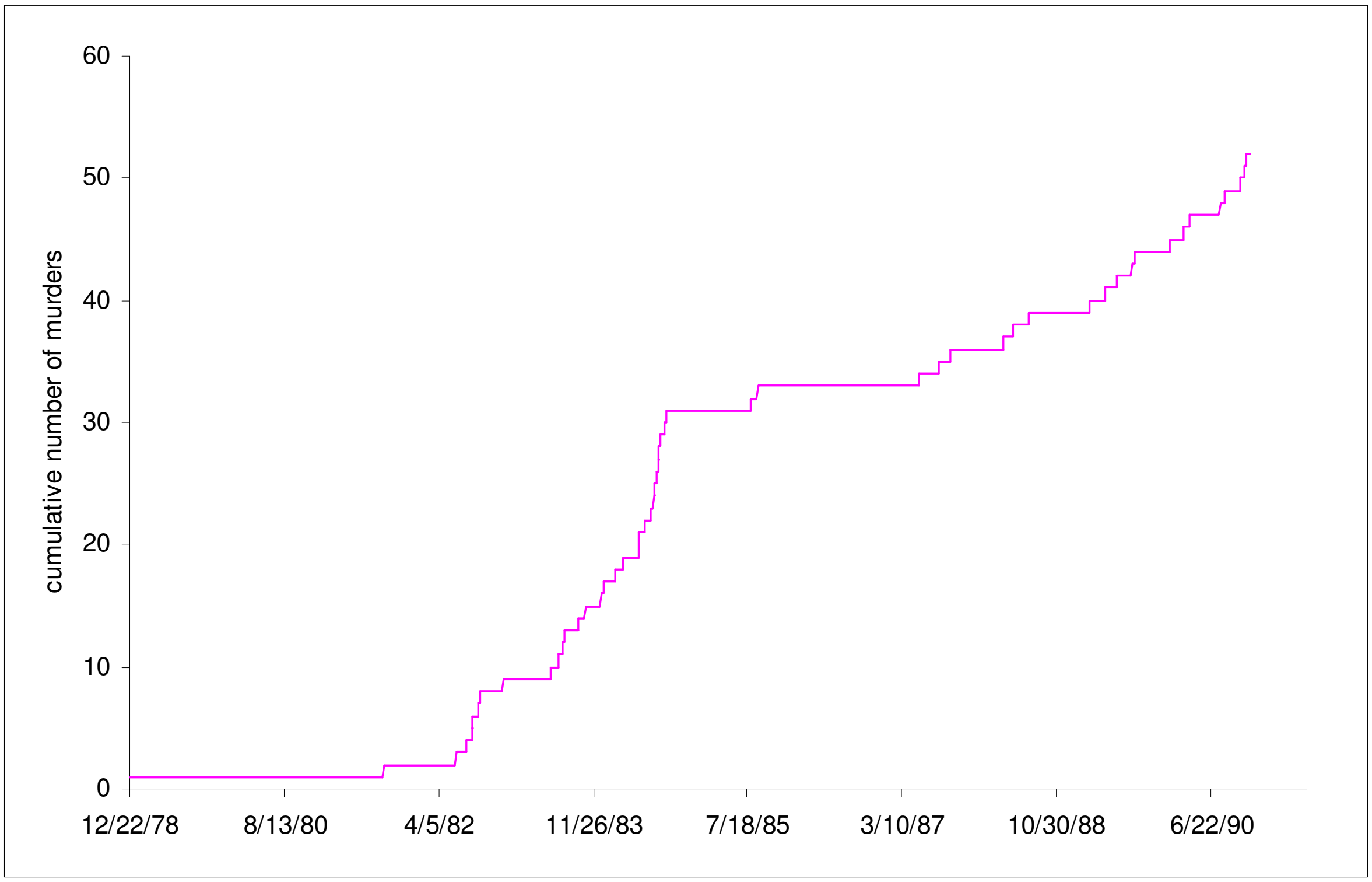

**Figure 1.** Chikatilo’s staircase shows how the total number of his murders grew with time. The time span begins with his first murder on 12/22/1978 and ends with his arrest on 10/20/1990. The shortest interval between murders was three days and the longest – 986 days. The murder dates were determined based on the date on disappearance of the person in question.

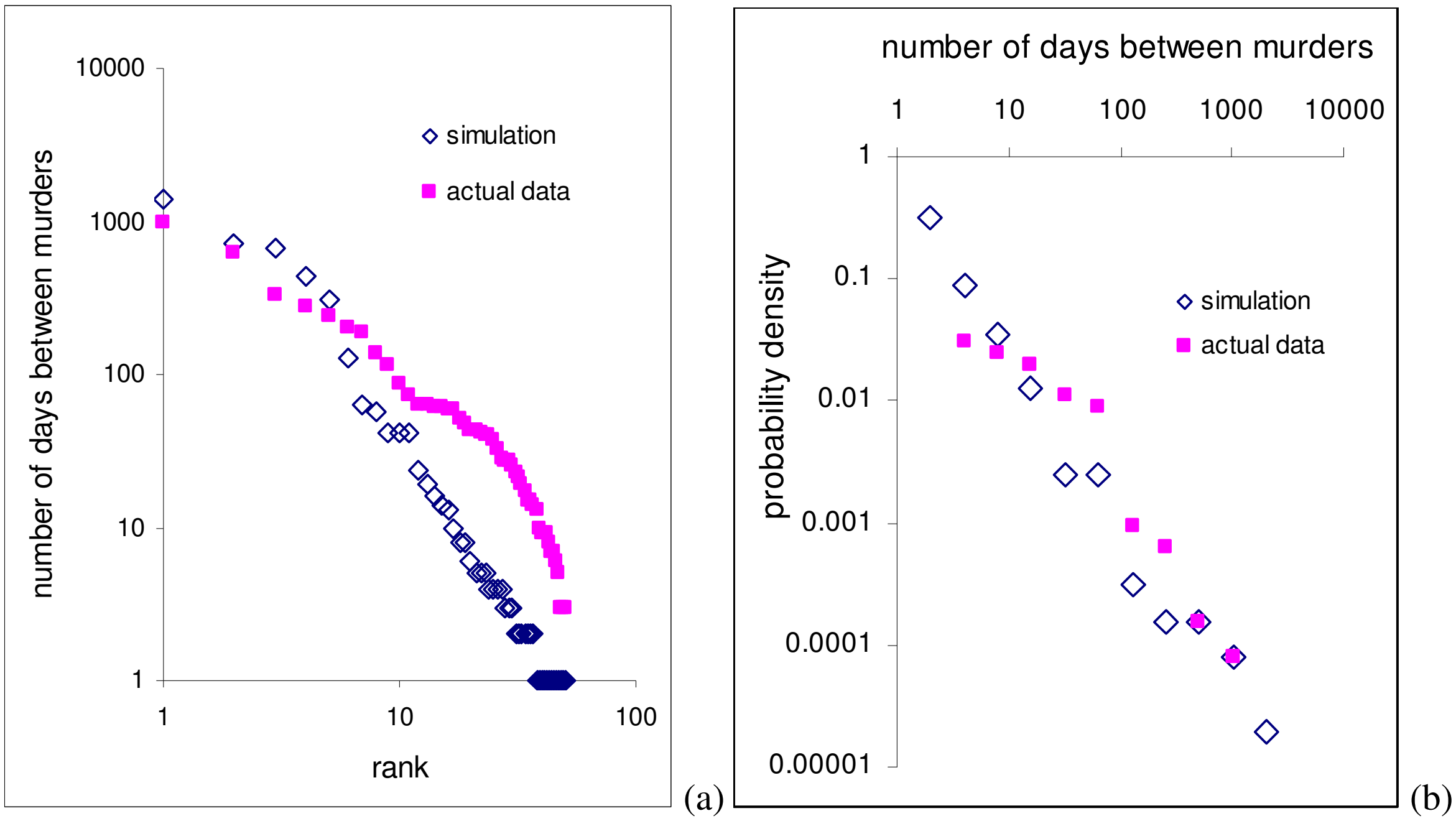

**Figure 2**. Distribution of step length (intervals between murders) in Zipfian (a) and probability density (b) representations.

Recently Osorio et al [3] reported a similar power-law distribution (with the exponent 1.5) of the intervals between epileptic seizures. Soon afterward they proposed [4] a self-organized critical model of epileptic seizures. They performed numerical simulations of their model and reproduced a power-law distribution of inter-seizure intervals. Almost simultaneously we proposed a stochastic neural network model of epileptic seizures [5], which was very similar to that of Osorio et al [4]. Unlike them, however, we solved our model analytically. Here we apply a similar model to explain the distribution of intervals between murders.

## The Model

We make a hypothesis that, similar to epileptic seizures, the condition, causing a serial killer to commit murder, arise from the simultaneous firing of large number of neurons in the brain. Our neural net model for epileptics [5] and serial killers is as follows. After a neuron has fired, it cannot fire again for a time interval known as the refractory period. Therefore, the minimum interval between the two subsequent firings of a neuron is the sum of spike duration and refractory period. This interval is few milliseconds and we will use it as our time unit. Consider one particular firing neuron. Its axon connects to synapses of thousands of other neurons. Some of them are almost ready to fire: their membrane potential is close to the firing threshold and the impulse from our neuron will be sufficient to surpass this threshold. These neurons will be firing at the next time step and they can be called "children" of our neuron in the language of the theory of branching processes [6]. Since the number of neurons connected to a given neuron is large and since each firing neuron will independently induce the firing of each of the neurons connected to it with a small probability, the number of firings induced by one firing neuron is binomially distributed with a large number of trials and a small success probability, which can be approximated by a Poisson random variable. In addition to induced firings, some neurons will fire spontaneously. We assume that the number of spontaneously firing neurons at each time step comes from a Poisson distribution with mean $p$.

Let us introduce the following random variables:

$X_n$ := number of firing neurons at time $n$
$Y_n$ := number of spontaneously firing neurons at time $n$
$Z_{n,j}$ := number of firings induced by the $j$th firing neuron at time $n$.

Then, the process ($X_n$) defines a discrete-time Markov chain which we obtain by assuming that the two collections of random variables $Y_n$ and $Z_{n,j}$ are collections of independent Poisson random variables with mean $p$ and $\lambda$, respectively, and by setting

$$X_{n+1} = Z_{n,1} + \cdots + Z_{n,Xn} + Y_{n+1}$$

The above equation can be rewritten as (here $E(\ldots)$ denotes the expectation value)

$$X_{n+1} = E(Z_{n,1} + \cdots + Z_{n,Xn}) + E(Y_{n+1}) + (Z_{n,1} + \cdots + Z_{n,Xn} - E(Z_{n,1} + \cdots + Z_{n,Xn})) + Y_{n+1} - E(Y_{n+1}) \quad (1)$$

In the limit of large $X_n$ we can apply the Central Limit Theorem and get

$$Z_{n,1} + \cdots + Z_{n,Xn} - E(Z_{n,1} + \cdots + Z_{n,Xn}) \Rightarrow X_n^{1/2} N(0, \lambda), \quad (2)$$

where $\Rightarrow$ means convergence in distribution and $N(0, \lambda)$is normal distribution with mean 0 and variance $\lambda$. By substituting Eq.(2) into Eq.(1) and noting that $E(Z_{n,1} + \cdots + Z_{n,Xn}) = \lambda X_n$ and $E(Y_{n+1}) = p$ we obtain that in the limit of large $X_n$

$$X_{n+1} = \lambda X_n + p + X_n^{1/2} z(0, \lambda), \quad (3)$$

where $z(0, \lambda)$ is a normal random variable with mean 0 and variance $\lambda$. We can rewrite Eq.(3) as

$$\Delta X = (\lambda - 1)X + p + \sqrt{X} z(0, \lambda) \quad (4)$$

The number of firing neurons, $X$, performs a random walk, with the size of the step proportional to $\sqrt{X}$. We can simplify Eq.(4) by changing variable from $X$ to $x = \sqrt{X}$. We get [5]:

$$\Delta x = \frac{(\lambda - 1)}{2} x - \left(\frac{1}{8} - \frac{p}{2}\right) \times \frac{1}{x} + \frac{1}{2} z \quad (5)$$

When $\lambda = 1$, the first term disappears. In the limit of large $x$, we can neglect the second term, since it is inversely proportional to $x$. Equation (5) reduces to $\Delta x = 1/2\, z$ which means that $\sqrt{X}$ performs a symmetric Gaussian random walk. A well known problem in random walk theory [7] is that of first return times. That is if the random walk is at point $x$, after how many steps it will return back to $x$ for the first time. A well known result is that the distribution of first return times follows a power law with an exponent of 3/2 [7]. We assume that the killer commits murder when the number of firing neurons reaches certain threshold. Then the distribution of first returns into murder zone (inter-murder intervals) is the same as the distribution of random walk's return times.

The model needs to be a bit more complex. We cannot expect that the killer commits murder right at the moment when neural excitation reaches a certain threshold. He needs time to plan and prepare his crime. So we assume that he commits a murder after the neural excitation was over the threshold for a certain period $d$. Another assumption that we make is that a murder exercises a sedative effect on the killer, causing the number of excited neurons to fall to the threshold. If we do not make this last assumption, the neural excitation will be in the murder zone for half of the time.

## Numerical Simulations

We made numerical simulations of the above model. We set $\lambda = 1$ which corresponds to the critical branching process in neuron firing. This selection is not arbitrary since some experiments [8] suggest that neural circuits operate in critical regime. There are also theoretical reasons to believe that the brain functions in a critical state [9]. The system was simulated for $2\times10^{11}$ time steps. Remember that time step is the sum of firing duration and refractory period.

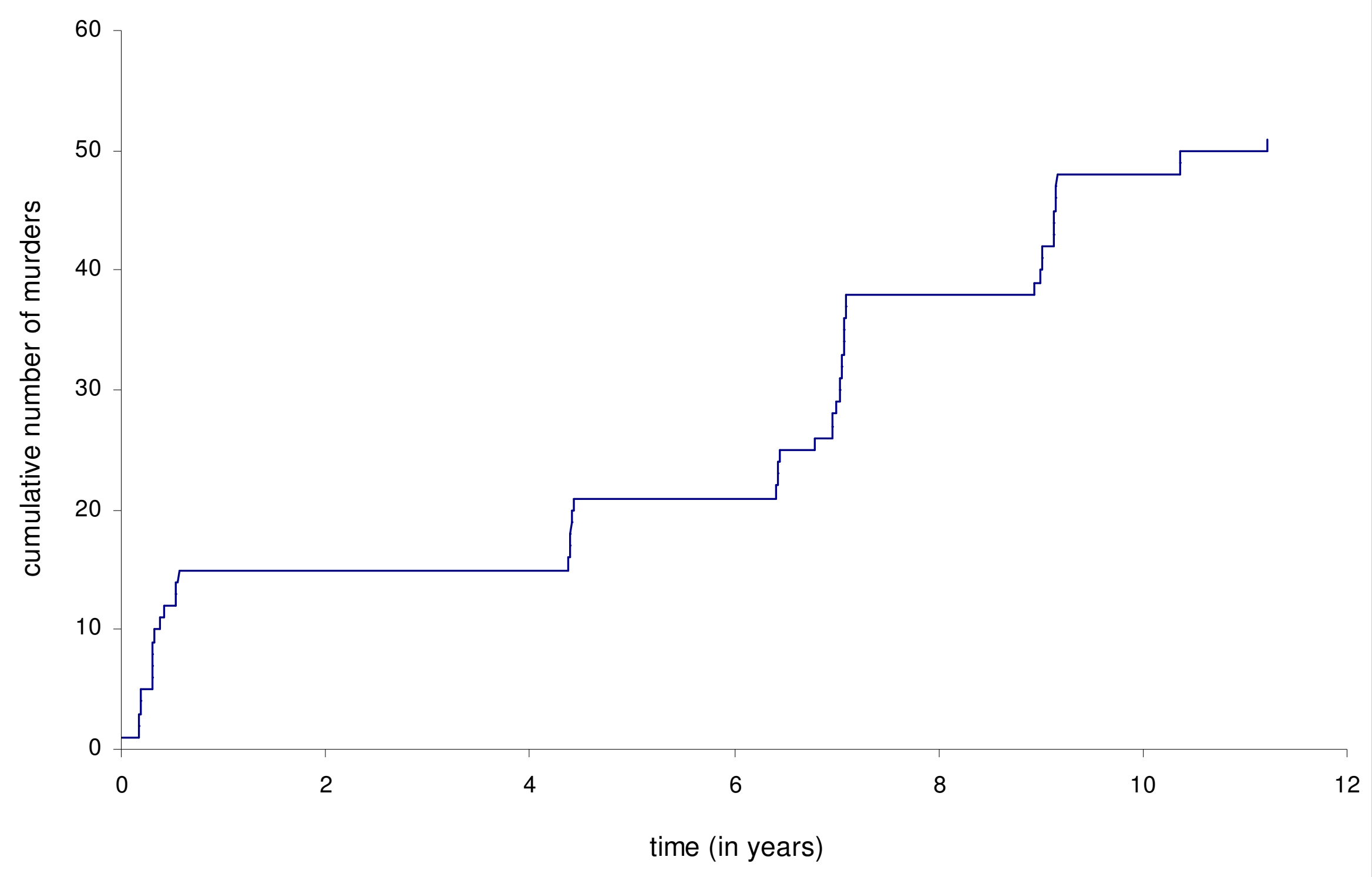

**Figure 3.** Results of numerical simulation of the stochastic serial killer model. The distribution of step length is shown in Figure 2.

A reasonable estimate for this is two milliseconds. Thus, our simulation run corresponds to about twelve years. The rate of spontaneous firing was set at $p = 0.1$. We set the intensity threshold at $10^9$ firing neurons. We set the time threshold, *d*, at 24 hours. Figures 2-3 show the results of these simulations. They qualitatively agree with the experimental data.

The major disagreement is probably that the actual minimum number of days between murders is three, while the simulation produces a dozen inter-murder periods of one day. One could enhance the model by introducing a murder success rate. That is, with certain probability everything goes well for the killer and he is able to commit the murder as he planned. If not, he repeats his attempt the next day. And so on. One could surely obtain a better agreement with experimental data, but this would be achieved by the price of introducing an extra parameter into the model.

## Discussion

To our knowledge, the only previous mathematical study of the time patterns of serial killers is that by Lange [10]. He used a polynomial map to express the ($n$+1)th inter-murder interval through $n$th and ($n$-1)th intervals. Afterward he used regression to find the coefficients of the polynomial that will maximize the correlation coefficient between the actual time series of inter-murder intervals and the time series obtained using the polynomial map. However, Lange did not propose any plausible mechanism to justify his model.

The random-walk models have already been used in Psychology to model human behavior [11]. However, those models deal with a choice between several opportunities, not with intervals between events. The connection between the present research and those earlier models may be worthy of investigation in the future.

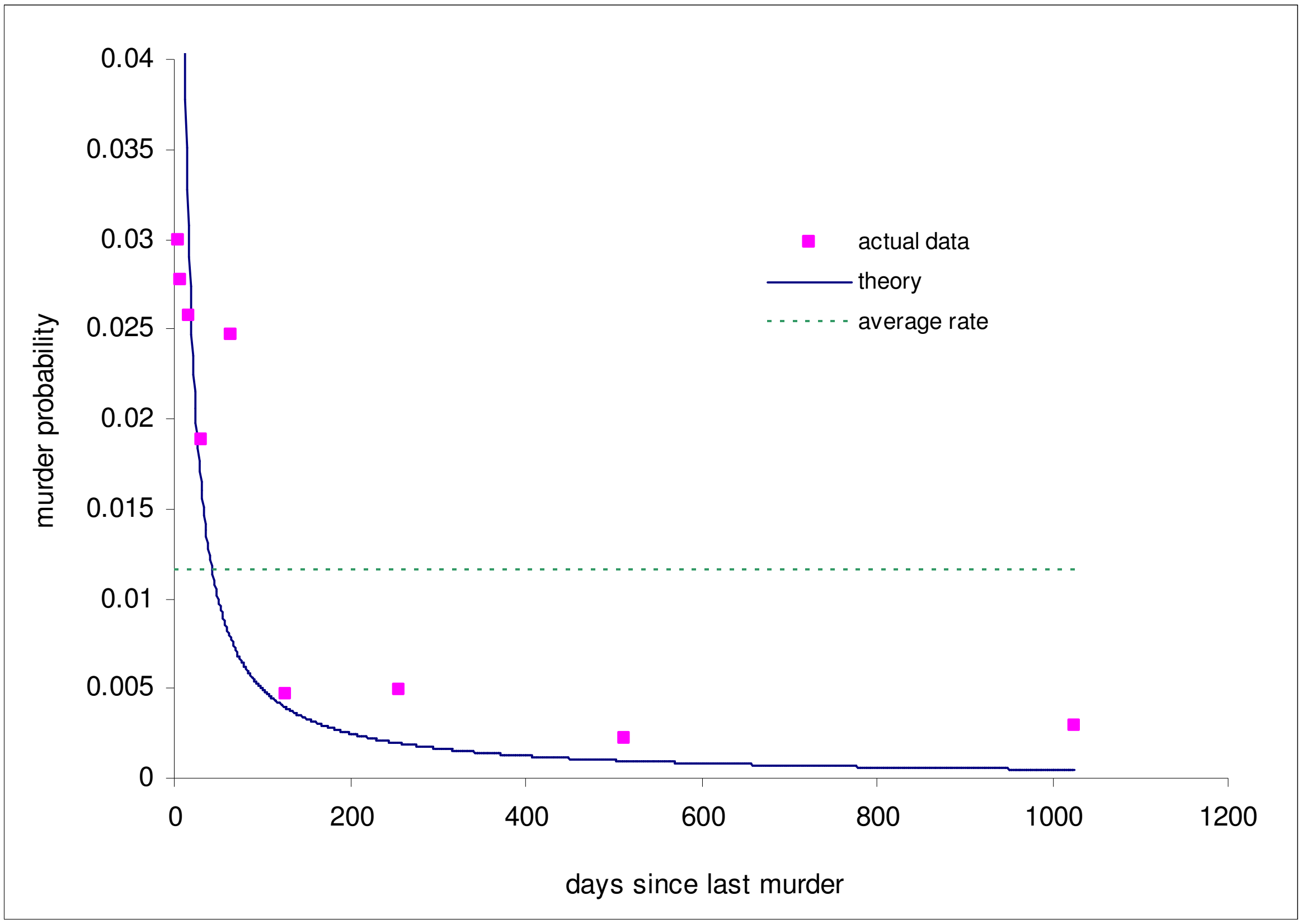

**Figure 4.** Daily murder probability as a function of the number of days passed since the last murder. The average murder rate is the total number of murders committed by Chikatilo divided by the length of the period during which he committed those murders.

An interesting question to ask is how the probability to commit a new murder depends on the time passed since the last murder. Suppose the killer committed his last murder $n$ days ago. From the random-walk approximation, we immediately get that the probability to commit a murder today is equal to $\frac{1}{n^{3/2}} \Big/ \int_n^\infty \frac{dm}{m^{3/2}} = \frac{1}{2n}$. Figure 4 shows this curve together with the actual data. The actual murder probability on the $n$th day after previous murder is computed the following way. The number of instances, when there had been no murder for $n-1$ days is the total number of murders that happened on $n$th or later day after the previous one. Thus the murder frequency on the $n$th day is the ratio of the number of murders, which happened exactly on *n*th day to the number of murders that happened on *n*th or later day. To estimate the probabilities we average these frequencies over corresponding bins. There is at least a qualitative agreement between theory and observation. In particular, the probability of a new murder is significantly higher than the average murder rate immediately after murder and is significantly lower than the average murder rate when long time has passed since the last murder.

Looking at Figure 2(b) one can notice a cusp near 100-day inter-murder interval. This may suggest that there is a characteristic time scale in the distribution. In one of their experiments with rats Osorio et al [4] have found a characteristic time scale mixed in with a scale-free behavior in the distribution of inter-seizure intervals. However in our case this is most likely a statistical fluctuation due to smallness of the sample.

Murder patterns of other serial killers are similar to that of Chikatilo. Figures 5 and 6 show the cumulative number of murders committed by Yang Xinhai [12] and Moses Sithole [13]. One obvious difference of Yang Xinhai's staircase from Chikatilo's staircase is the higher step heights: Yang Xinhai killed up to five people at a time. The distribution of step length, however, is very similar as one can see from Figure 7. One noticeable difference with the Chikatilo case is that we see two-day intervals between murders for Yang Xinhai and one-day intervals for Moses Sithole. This, however, does not imply any real difference in murder patterns but is most likely due to chance: no killer committed enough murders to accumulate accurate intervals statistics. This is also obvious from how scattered the plots in Figure 2 and 7 look. We can partially overcome this problem by combining intervals from all three killers. Figure 8 shows such plot. As one can see, it is far less scattered than the other plots, although a strong deviation from a power law for short inter-murder intervals remains.

This work was supported in part by the NIH grant No. R01 GM105033-01. Computations were performed on UCLA IDRE Hoffman2 cluster.

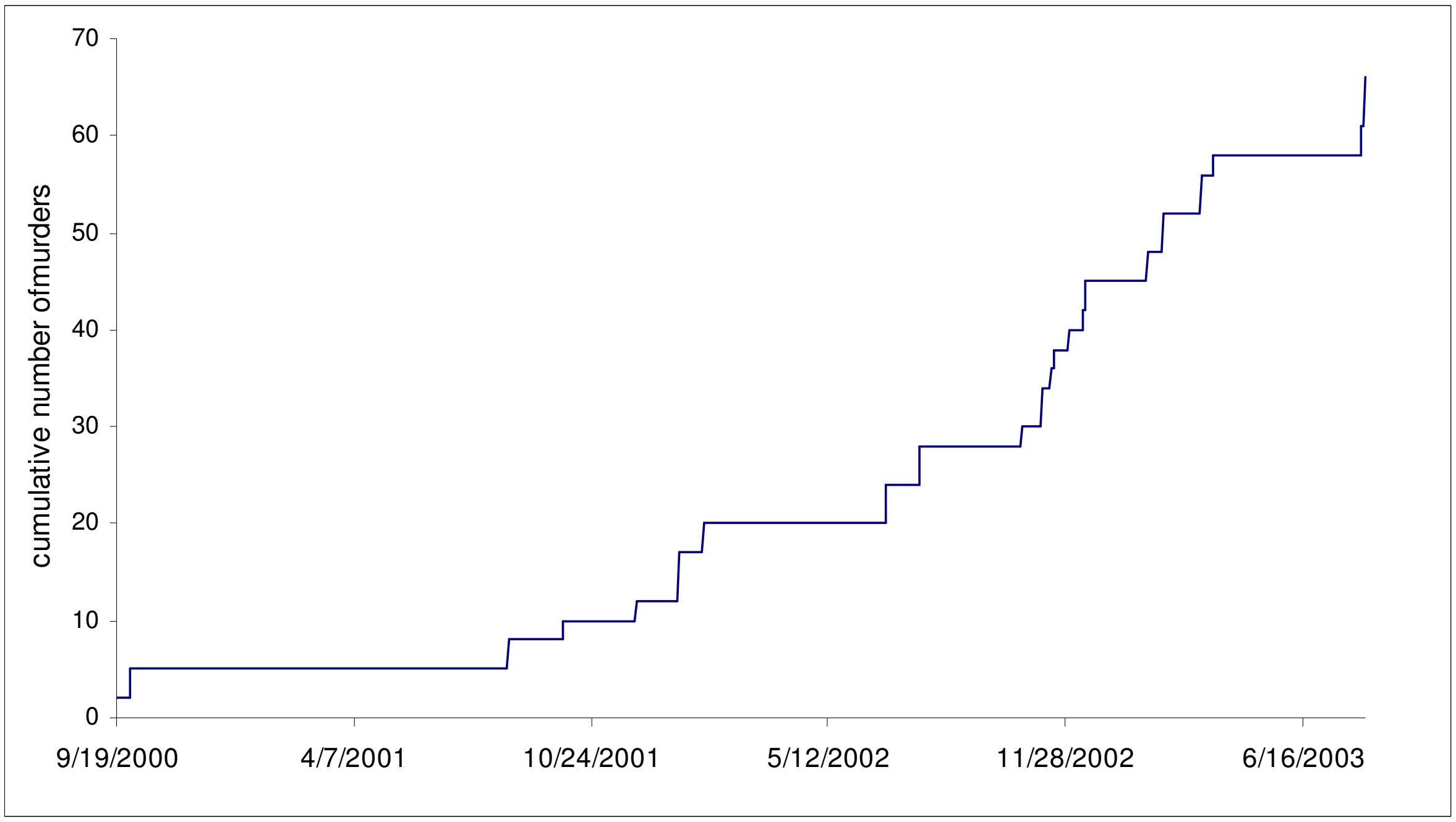

**Figure 5.** Cumulative number of murders committed by Yang Xinhai.

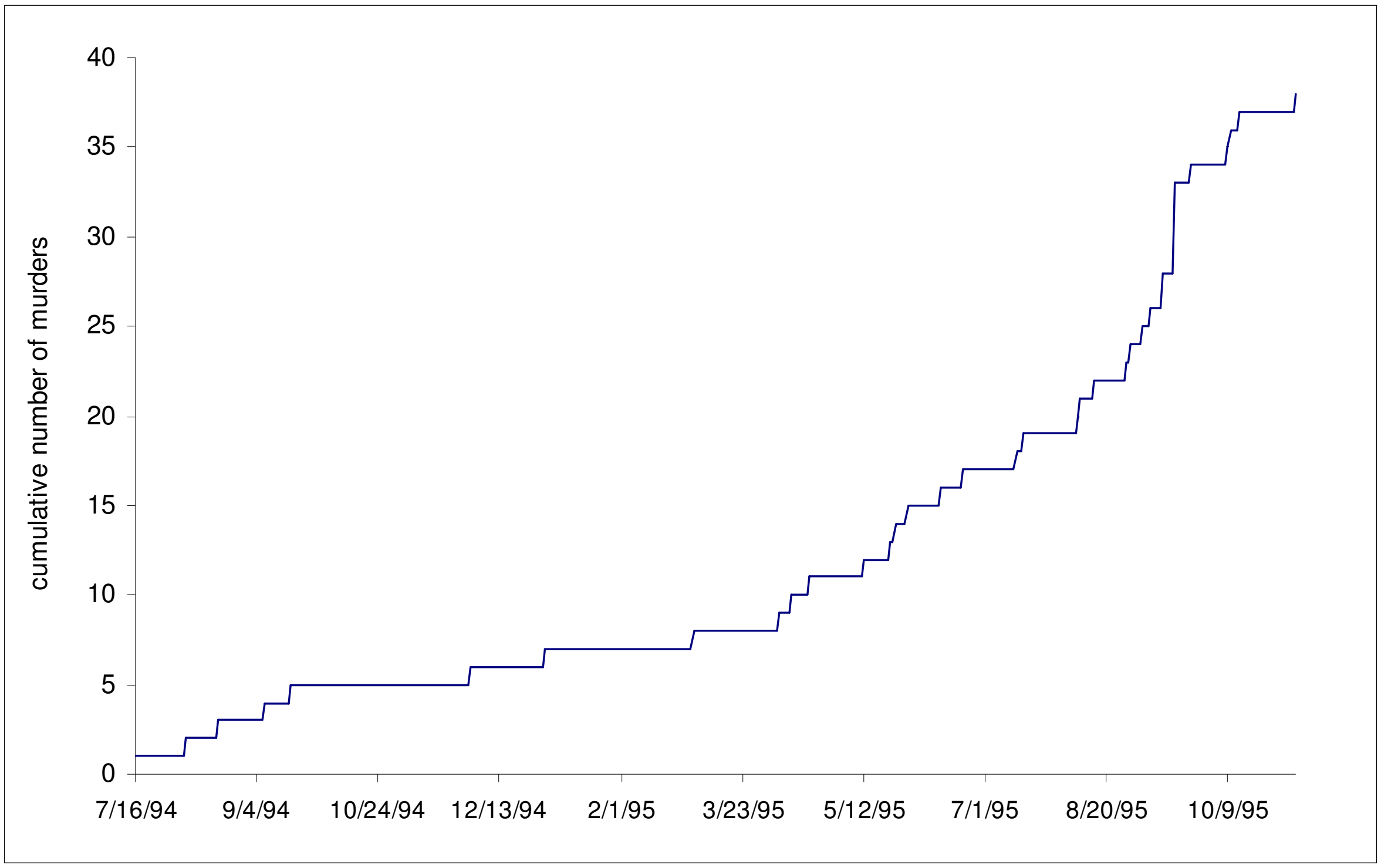

**Figure 6.** Cumulative number of murders committed by Moses Sithole.

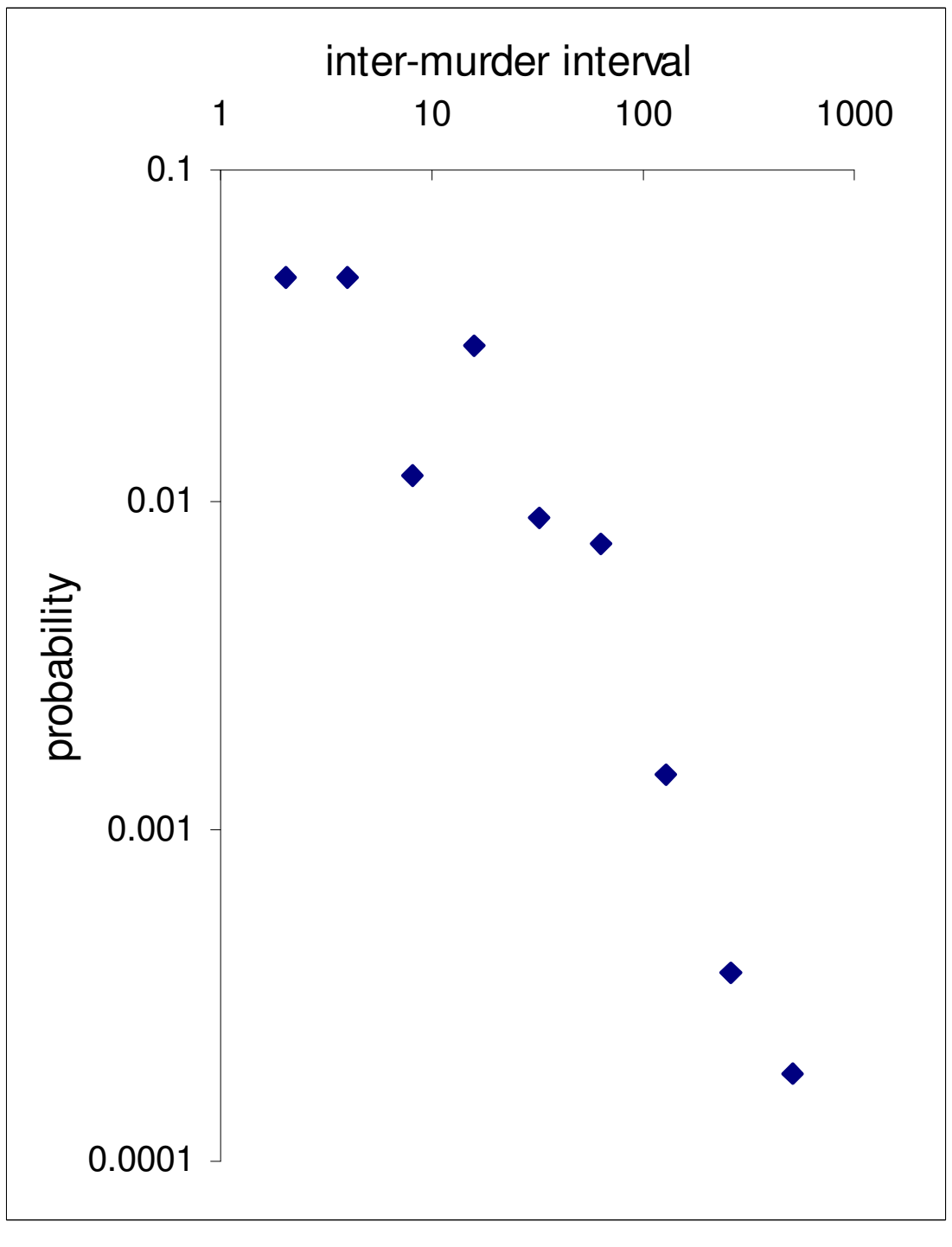

(a)

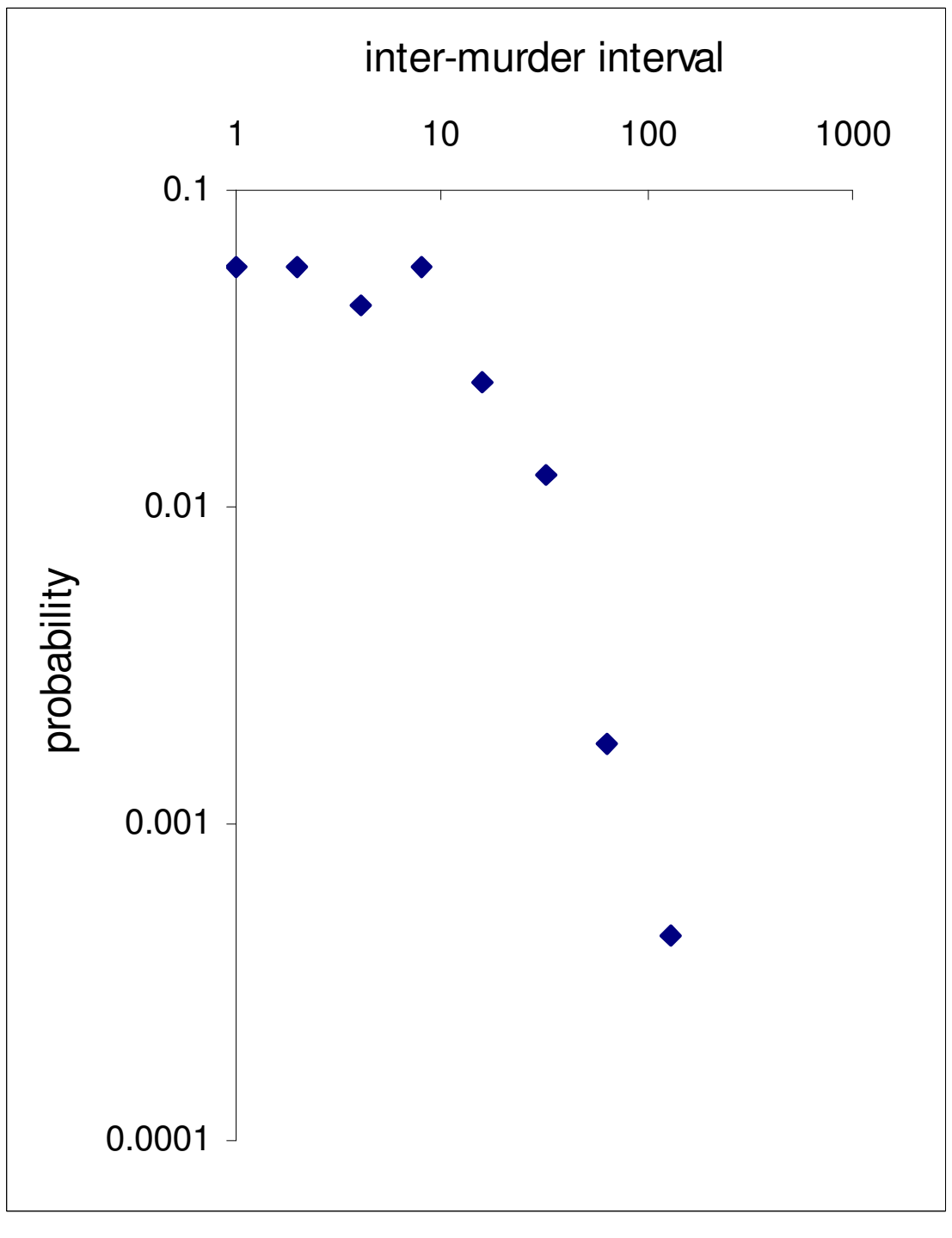

(b)

**Figure 7.** Probability distribution of inter-murder intervals for Yang Xinhai (a) and Moses Sithole (b).

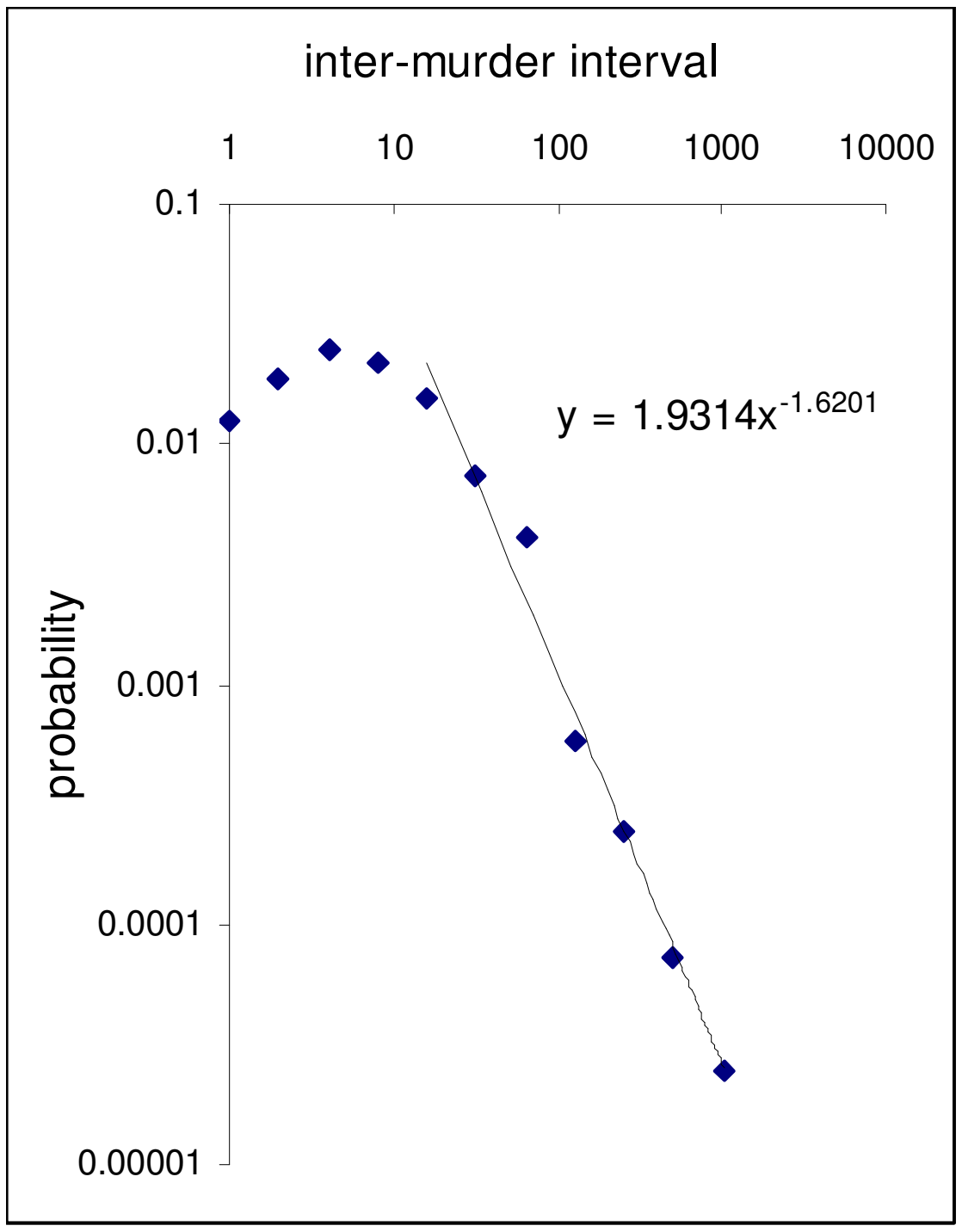

**Figure 8.** Probability distribution of combined inter-murder intervals for Andrei Chikatilo, Yang Xinhai, and Moses Sithole.